\documentclass[a4paper,11pt]{article}
\pdfoutput=1 % if your are submitting a pdflatex (i.e. if you have
\usepackage{jinstpub} % for details on the use of the package, please
\usepackage{graphicx}
\usepackage{subfigure}
\usepackage{float}

\title{\boldmath A low-energy sensitive compact gamma-ray detector based on $LaBr_3$ and SiPM for GECAM}

\author[a,b,c]{P Lv,}
\author[a]{S.L Xiong,}
\author[a,b,1]{X.L Sun,\note{Corresponding author.}}
\author[a,b]{J.G Lv,}
\author[a]{Y.G Li}

\affiliation[a]{Institute of High Energy Physics, CAS,\\Beijing 100049, China}
\affiliation[b]{State Key Laboratory of Particle Detection and Electronics, Institute of High Energy Physics, CAS,\\Beijing 100049, China}
\affiliation[c]{University of Chinese Academy of Sciences,\\Beijing 100049, China}

\emailAdd{sunxl@ihep.ac.cn}

\abstract{The Gravitational wave high-energy Electromagnetic Counterpart All-sky Monitor (GECAM) project is the planned Chinese space telescope for detecting the X- and gamma-ray. It consists of two micro-satellites in low earth orbit with the advantages of instantaneous full-sky coverage. Due to the limitation of size, weight and power consumption of micro-satellites, silicon photomultipliers (SiPMs) are used to replace the photomultiplier tubes (PMTs) to assemble a novel gamma-ray detector. A prototype of a SiPM array with $LaBr_3$ crystal is built and tested, and it shows a high detection efficiency ($70\%$ at 5.9~keV) and an acceptable uniformity. The low energy X-ray of 5.9~keV can be detected by single channel readout circuit, and the energy resolution are $125\%$ (FWHM) at 5.9~keV, $6.5\%$ at 662~keV and $3.5\%$ at 1332~keV. The design and the performance of the detector are discussed in this paper.}

\keywords{electromagnetic counterpart, gamma-ray detector, $LaBr_3$ crystal, SiPM, low threshold, energy resolution}

\proceeding{N$^{\text{th}}$ Workshop on X\\
  when\\
  where}

\begin{document}
\maketitle
\flushbottom

\section{Introduction}
\label{sec:intro}

First gravitational wave event GW150914~\cite{PhysRevLett.116.131103,PhysRevLett.116.131102, PhysRevLett.116.061102}  and double neutron star merger event GW170817~\cite{PhysRevLett.119.161101}, heralded the new era of gravitational wave multi-messenger astronomy. The energy region from keV to MeV is an important window for electromagnetic counterpart research. The two micro-satellites of GECAM, which are in the same low earth orbit with opposite orbital phase for full-sky coverage, are planned to launch in 2020. Each satellite consists of 25 gamma-ray detectors (GRD) and 8 charged-particle detectors (CPD). As the main instrument, GRD is used to monitor electromagnetic counterparts of gravitational wave events from 6~keV to 2~MeV.
\par\indent
Traditional PMTs are bulky and high power consumption, which make them inappropriate for space, especially in micro-satellites. SiPM, a novel silicon-based photodetector, has processed attractive capabilities such as a large dynamic range, single-photon sensitivity, high photon detection efficiency (PDE), insusceptibility to magnetic fields and low power consumption~\cite{1748-0221-10-08-C08017}. However, the high thermal-noise~\cite{SenSLdatasheet} is an obstacle to detect low-energy rays especially below 10~keV at room temperature. Hence the high light yield and fast decay time of the scintillator are the key factors to improve signal-noise-ratio (SNR) for low energy X-rays detection. The $LaBr_3$ crystal is a good candidate because of its fast decay time (16~ns), high light yield ($>$60,000~photons/MeV)~\cite{1621373, 1310614, QUARATI2007115}, high energy resolution and good linearity. A prototype for GECAM is designed with a $LaBr_3$ crystal (76.2~mm in diameter) and a SiPM array ($50.44\times50.44~{{mm}^2}$). 
\par\indent
The experiment results show that the 5.9~keV full-energy peak of the ${}^{55}Fe$ source can be detected with a $70\%$ efficiency. A simple single-channel readout was used. Thermal noise was suppressed by using a proper discrimination level. This design provides a new solution for space gamma-ray detectors in the future.

\section{The GRD Structure}
\begin{figure}[H]
	\centering
	\includegraphics[width=9cm]{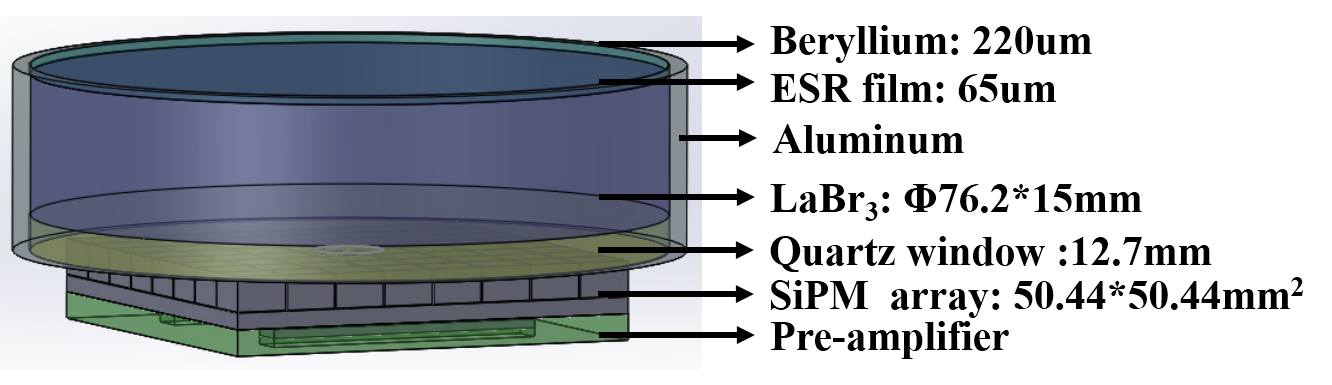}
	\caption{The diagram of the GRD.}
	\label{Detector}
\end{figure}

\begin{figure}[H]
	\centering
	\subfigure[]{
		\begin{minipage}[b]{0.3\textwidth}
			\centering
			\includegraphics[width=2.5cm]{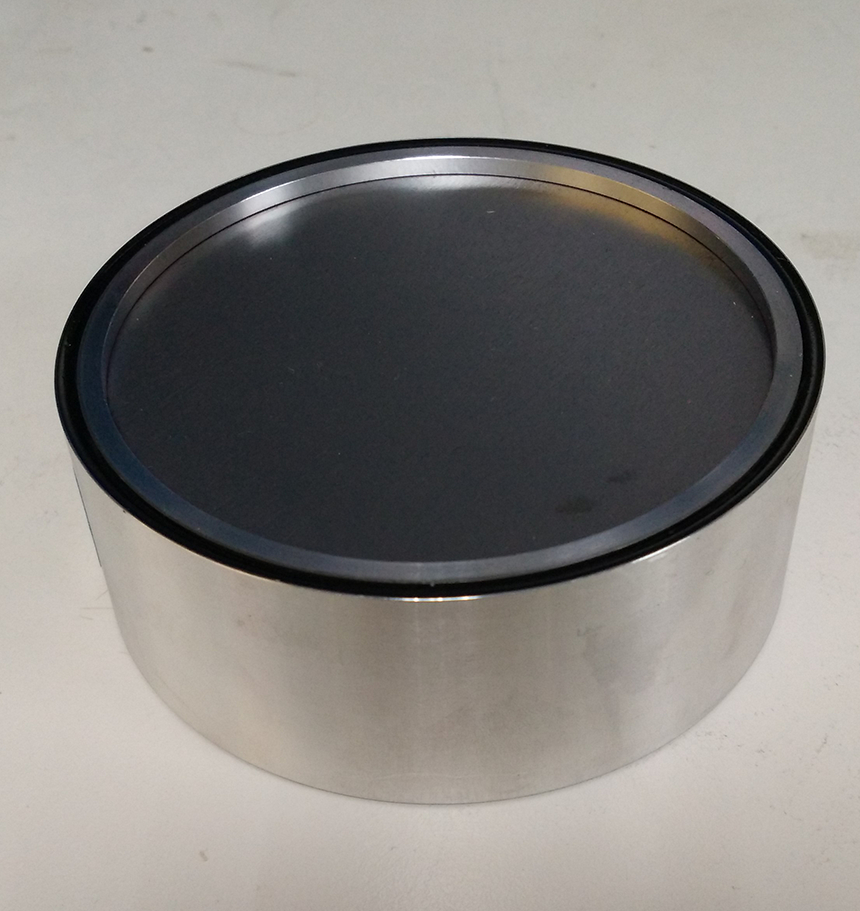}
			\label{crystal}
		\end{minipage}
	}
	\subfigure[]{
		\begin{minipage}[b]{0.3\textwidth}
			\centering
			\includegraphics[width=3.6cm]{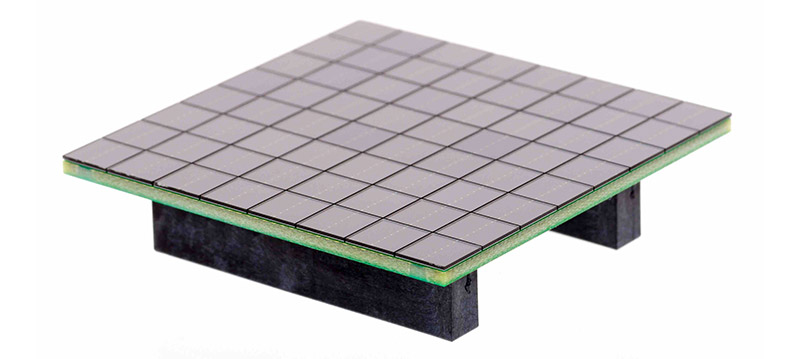}
			\label{sipm}
		\end{minipage}
	}
	\subfigure[]{
		\begin{minipage}[b]{0.3\textwidth}
			\centering
			\includegraphics[width=2cm]{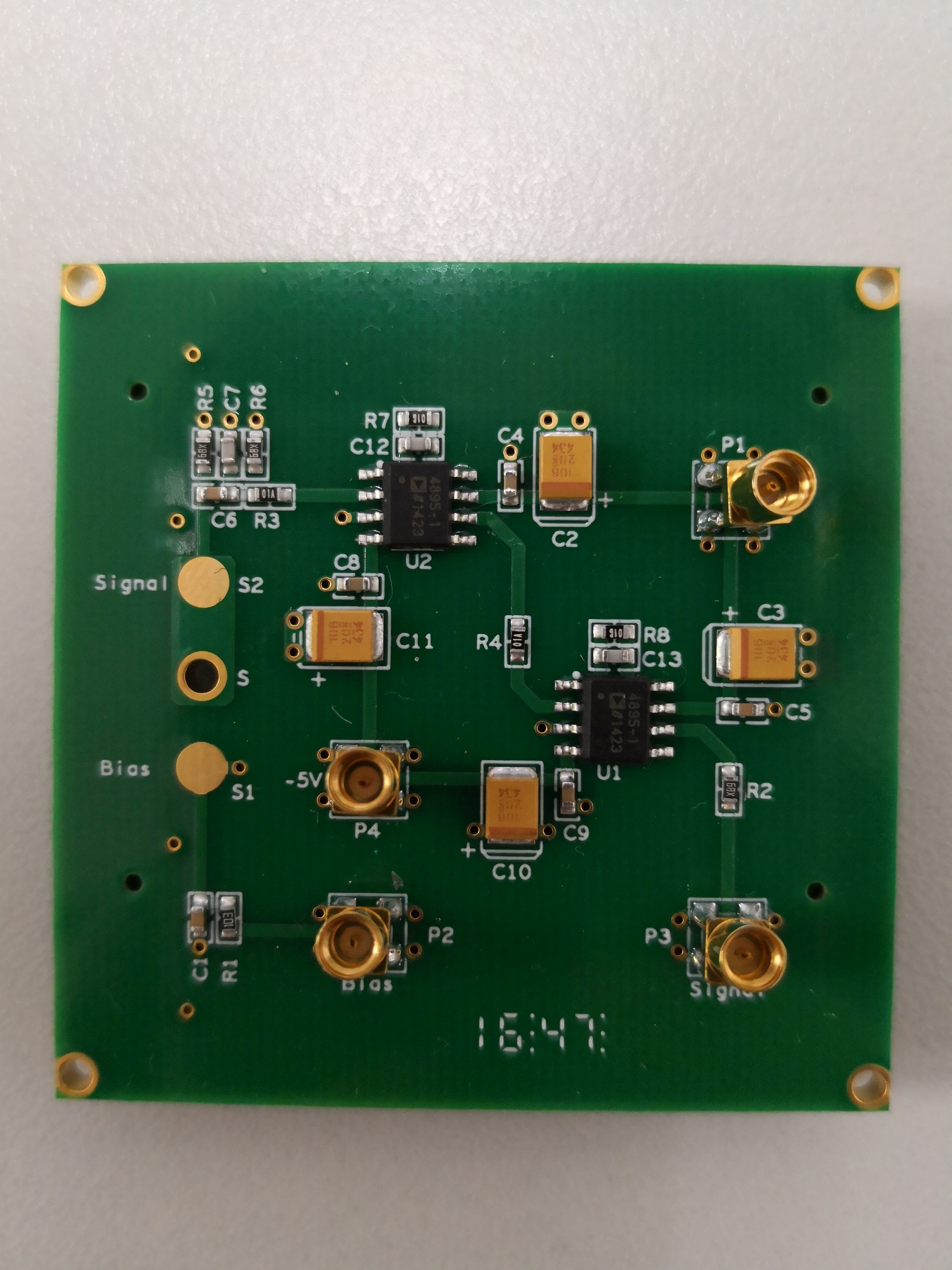}
			\label{board}
		\end{minipage}
	}
	\caption{The $LaBr_3$ crystal, the SiPM array and the pre-amplifier.}
\end{figure}

%\label{sec:intro}
The schematic diagram of the GRD is shown in Fig.\ref{Detector}.
The $LaBr_3$ crystal was grown by Saint Gobain Crystal Company~\cite{Saint_Gobain}. It is 76.2~mm in diameter and 15~mm thick (Fig\ref{crystal}), which provides $35\%$ detection efficiency for 1~MeV gamma-rays. The crystal is sealed in an aluminium enclosure with a quartz window. A 220~$\mu$m beryllium and 65~$\mu$m enhanced specular reflector (ESR) film are used as the entrance window.  The optical signals are collected by the SiPM after passing through the quartz window and then converted into electrical signals. The electrical signals are amplified by a pre-amplifier before being recorded.
\par\indent
The studied SiPM devices are of type ArrayJ-60035-64P from SensL, with 6mm$\times$6mm active area each pad and 35~$\mu$m pixel, resulting in a total of 22292 pixels. There are 64 pads (50.44mm$\times$50.44mm active area) in each SiPM array. The SiPMs are operated at the bias voltage of 5~V above the breakdowm voltage, $V_{bd}$ +5~V; their photon detection efficiency (PDE) is $51\%$ at the peak wavelength of 420~nm. The $LaBr_3$ has an emission spectrum with a peak at 380~nm, which is close to PDE peak of SiPM.
\par\indent
The pre-amplifier (Fig.\ref{board}) is home made, it has a gain of 25 and low power consumption. A filter circuit is designed to keep the bias for the SiPM array stable. 
The gap between the quartz window and the array is filled with silicone oil to increase light transmittance. Moreover, the area of quartz window that does not touch the array is covered by Teflon film to enhance light reflectance. The GRD is placed in a dark box during measurments. Single-channel electronics is used to read the total 64 pads SiPM array.

\section{The Test System}
\begin{figure}[H]
	\centering
	\includegraphics[width=0.9\textwidth]{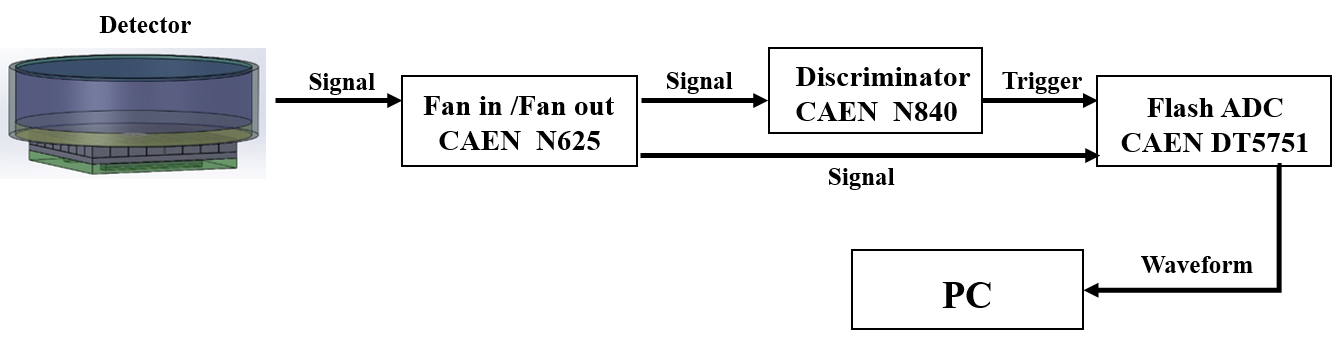}
	\caption{The block diagram of the test system.}
	\label{chart}
\end{figure}

The test system works at room temperature, and the diagram is shown in Fig.\ref{chart}. CAEN N625 is used to fan-out electral signals: one channel enters CAEN DT5751 digitizer directly and then is recorded by the PC, while the other generates a trigger by the discriminator CAEN N840. The threshold is 7~mV, which is about twice of the baseline noise. The time window of each waveform is $3~\mathrm{\mu s}$. The bias of SiPM is $V_{bd}$ +4V to obtain a proper gain ($>10^6$). The power supply for the preamplifier is $\pm$ 5V with 0.06~W power consumption, being the total consumption for each GRD less than 0.1~W.
\par\indent
A series of calibraion radioactive sources with energy ranging from 5.9~keV to 1332~keV are used to study the detector performance. The internal radioactivity of the $LaBr_3$ crystal has two low-energy X-rays, 5.6~keV and 37.4~keV~\cite{QUARATI201246}. The list of the sources with the energies of their emitted photons are shown in Table.\ref{tab:1}.
\begin{table}[H]
	\centering
	\caption{\label{tab:1} The radioactive sources and activity.}
	\smallskip
	\begin{tabular}{|c|l|l|l|}
		\hline
		Source&Energy~(keV)&Branching ratio (\%)&Activity~(Bq)\\
		\hline	
		${}^{138}La$    &5.6,  37.4   &--		&--   \\
		${}^{55}Fe$      &5.9         &28    &9058	\\
		${}^{241}Am$  &59.5     	&35.9	&2.6E+05	\\
		${}^{57}Co$	  &122,  136    &85.6, 10.68	&7479, 933	\\
		${}^{133}Ba$  &81,  356     &34.06, 62.05		&9.7E+04, 1.8E+05	\\
		${}^{137}Cs$	  &662    &85.1		&1.1E+06	\\
		${}^{60}Co$  &1173,  1332     	&99.97, 99.98		&1707, 1707	\\
		\hline
	\end{tabular}
\end{table}
\section{The Detector Performance}
\subsection{The low-energy performance}
\begin{figure}[H]
	\centering
	\includegraphics[width=0.9\textwidth]{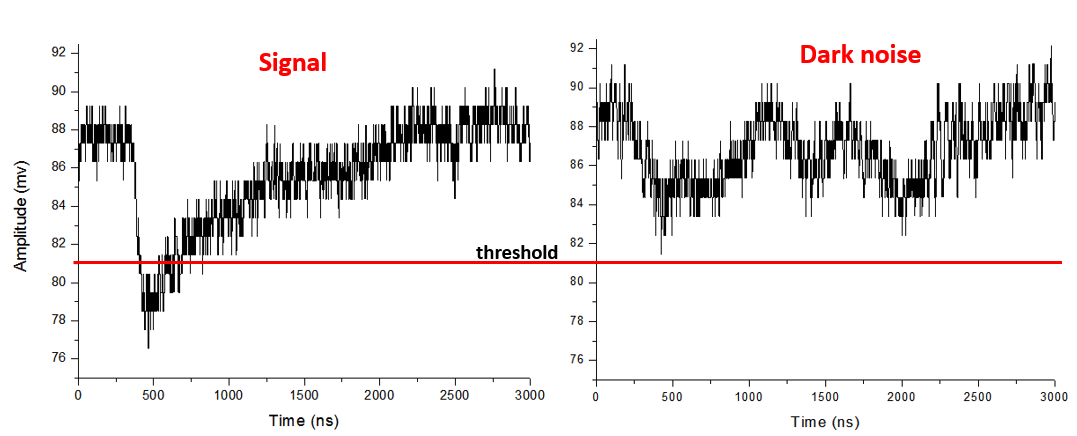}
	\caption{The typical waveform of 5.9 keV X-rays and the dark noise at room temperature.}
	\label{sigandnoi}
\end{figure}

\begin{figure}[H]
	\centering
	\includegraphics[width=6.5cm]{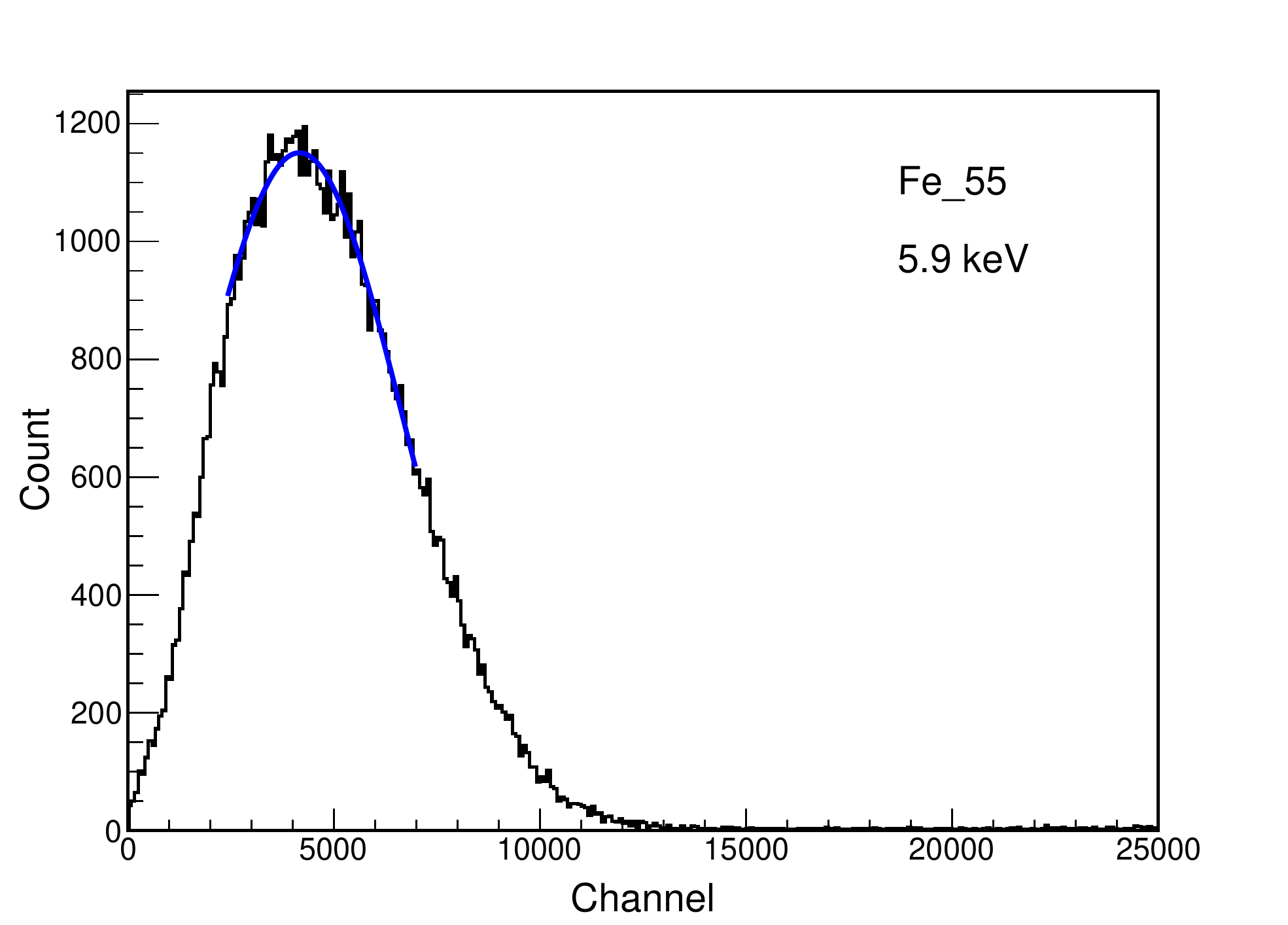}
	\includegraphics[width=6.5cm]{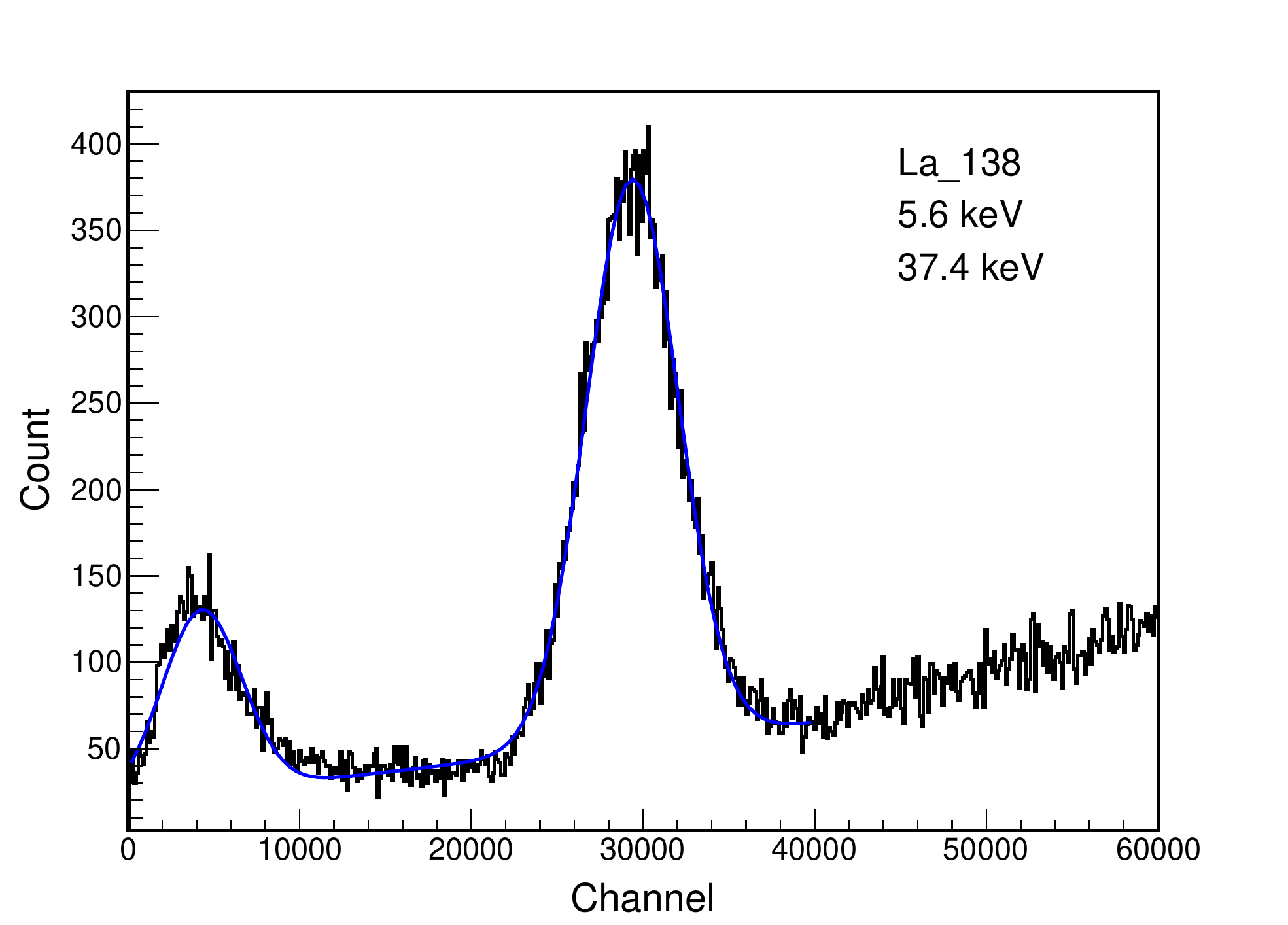}
	\caption{The spectra of ${}^{55}Fe$ and  ${}^{138}La$ radioactive sources. The energy resolutions are $95.6\%$ at 5.9~keV, $126.6\%$ at 5.6~keV and $24.1\%$ at 37.4~keV.}
	\label{lowenergy}
\end{figure}

The 5.9~keV and the 5.6~keV X-rays are used to study the GRD low-energy performance. Fig.\ref{sigandnoi} depicts the typical waveform of 5.9~keV X-ray and SiPM dark noise at room temperature. As the red line shows, a threshold at 7~mV can distinguish the signals and the noise. The energy spectra of 5.9~keV, 5.6~keV and 37.4~keV are shown in Fig.\ref{lowenergy}, the full-energy peaks are well seperated from the background being the energy resolution $95.6\%$, $126.6\%$ and $21.4\%$, respectively. This result indicates that the GRD is sensitive to the X-rays as low as 6~keV, which satisfies GECAM requirement.
\par\indent
The detection efficiency for low energy X-rays has been preliminary explored by measuring the event rate of an activity-known ${}^{55}Fe$ source. Table\ref{tab:2} presents essential information on the iron source. The source is deposited on a metal sheet which is placed in contact with the beryllium window during the measurements. The gap between the source and the crystal is just 285~$\mu$m due to the 220~$\mu$m beryllium window and the 65~$\mu$m ESR reflective film, hence the incident angle of the source to the crystal is about 2$\pi$. The activity of 5.9~keV X-rays can be calculated according to the half-life equation, and it was about 2536~Bq during the experiment. The event rate should be about 1268~Hz with $100\%$ efficiency, however, the measured  event rate is 890~Hz, for which the background events (280~Hz) have been deducted. The detection efficiency is equal to the ratio of measure events to the calculated activity, which is about $70\%$ (890/1268) for our GRD, the lost part being due to absorption in the beryllium window and the ESR reflective film. This efficiency satisfies the target of GRD (>$50\%$ at 6~keV). Of course, considering the obliquely incident attenuation of the thickness of the beryllium sheet, the incident angle should be less than $2\pi$. Therefore, the measured efficiency is a relatively rough estimate. We plan to achieve better results by using a precise X-ray apparatus
in the future.
\begin{table}[H]
	\centering
	\caption{\label{tab:2} The information of ${}^{55}Fe$ source.}
	\smallskip
	\begin{tabular}{|c|c|c|c|c|c}
		\hline
		Source&Energy~(keV)&Branching ratio for 5.9 keV ($\%$)&$T_0$~(year)&$T_0  
		$ activity~(Bq)\\        
		\hline
		${}^{55}Fe$  &5.9   &28    &1986   &2.59E07     \\
		\hline
	\end{tabular}
	\label{sourceinformation}
\end{table}
\subsection{Energy Resolution}
\begin{figure}[H]
	\centering
	\includegraphics[width=6.5cm]{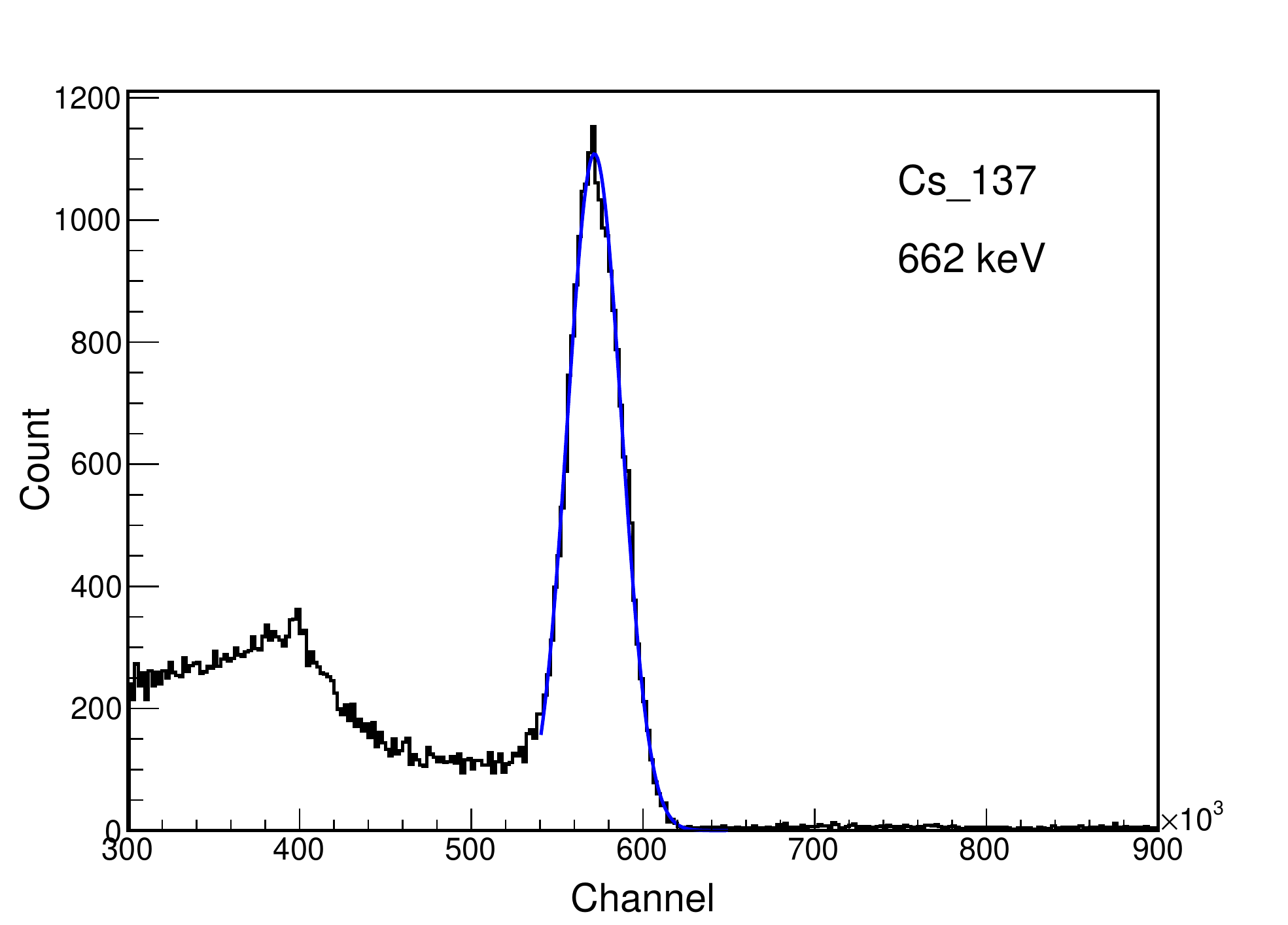}
	\includegraphics[width=6.5cm]{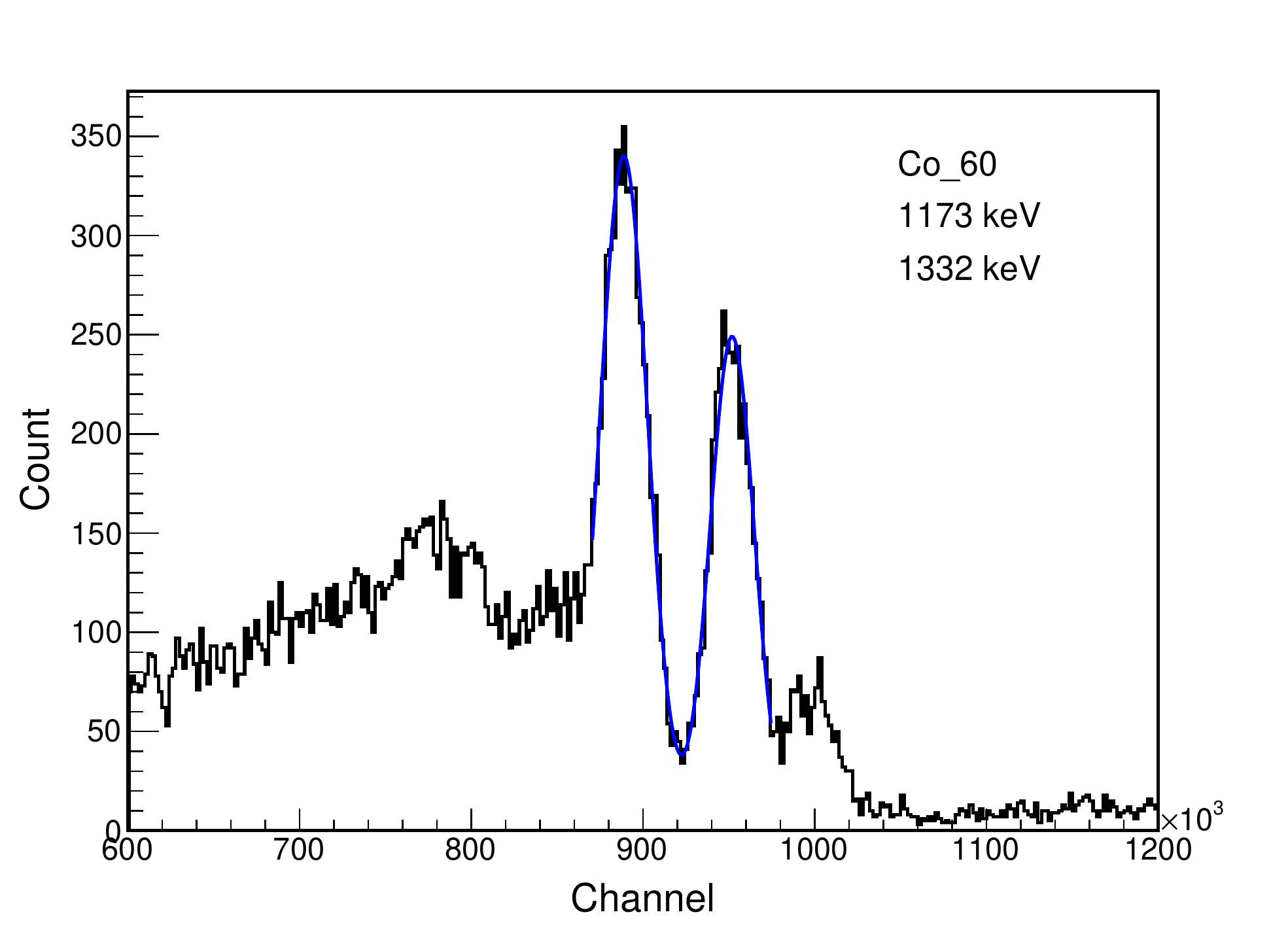}
	\caption{The spectra of ${}^{137}Cs$ and  ${}^{60}Co$, The energy resolutions are $6.5\%$ at 662~keV and $3.38\%$ at 1332~keV.}
	\label{enereo}
\end{figure}

\begin{figure}[H]
	\centering
	\includegraphics[width=12cm]{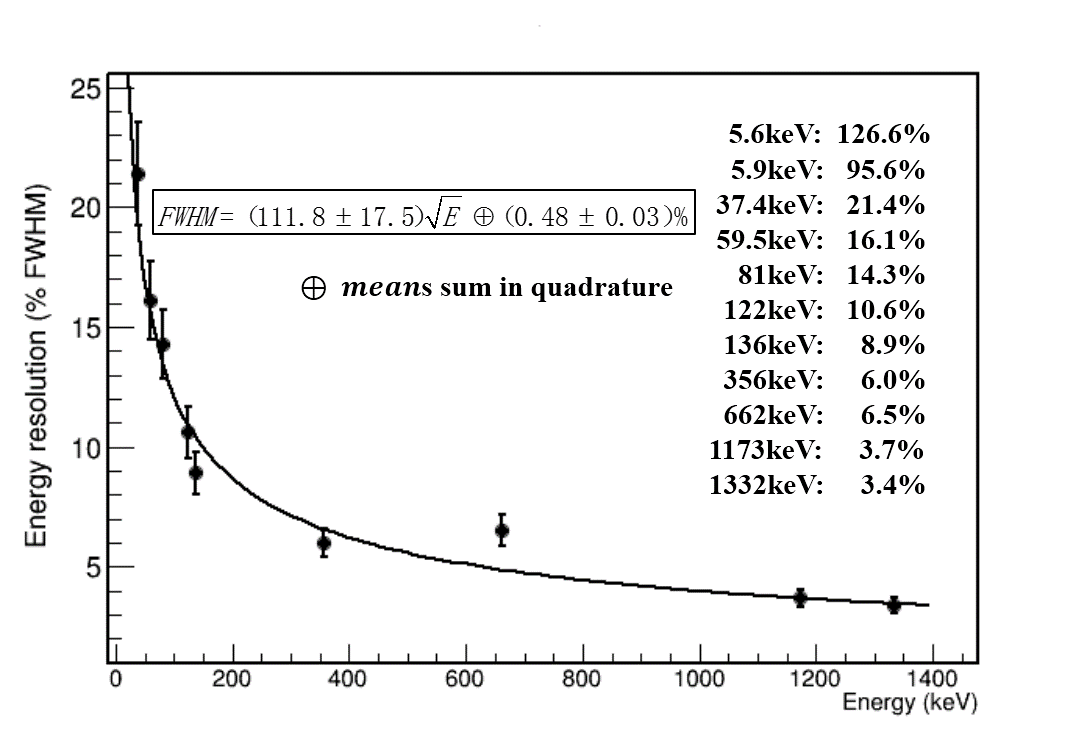}
	\caption{The plot of energy resolution versus energy and the datasheet.}
	\label{resolution}
\end{figure}

Scintillation detectors typically use 662~keV and 1332~keV to characterize their energy resolution. Fig.\ref{enereo} shows the energy spectra of ${}^{137}Cs$ and ${}^{60}Co$ sources, the energy resolutions are $6.5\%$ at 662~keV and $3.38\%$ at 1332~keV. The plot of energy resolution versus energy is shown in Fig.\ref{resolution}. A function FWHM($\%$) = $a/(E) ^{1/2}$$\bigoplus$b$\%$ ~\cite{fwhm1} was used to fit the data on energy resolution, $\bigoplus$ means sum in quadrature. This can be
compared with measurements from other groups with a small
$LaBr_3$ crystal coupled to a PMT, in which resolution of $3\%$
and $2\%$ was obtained for 662~keV and 1332~keV gamma-rays, respectively~\cite{QUARATI2007115}. In the present work, a larger crystal was used
read by SiPMs. The energy resolution is affected by dark noise. For SiPM, the dark noise is 30-70~kHz/$mm^2$ at room temperature~\cite{SenSLdatasheet}, which is much higher than PMT. Both cross-talk and after-pulse will deteriorate the energy resolution~\cite{1748-0221-11-01-C01078}.

\subsection{Linearity}
\begin{figure}[H]
	\centering
	\includegraphics[width=12cm]{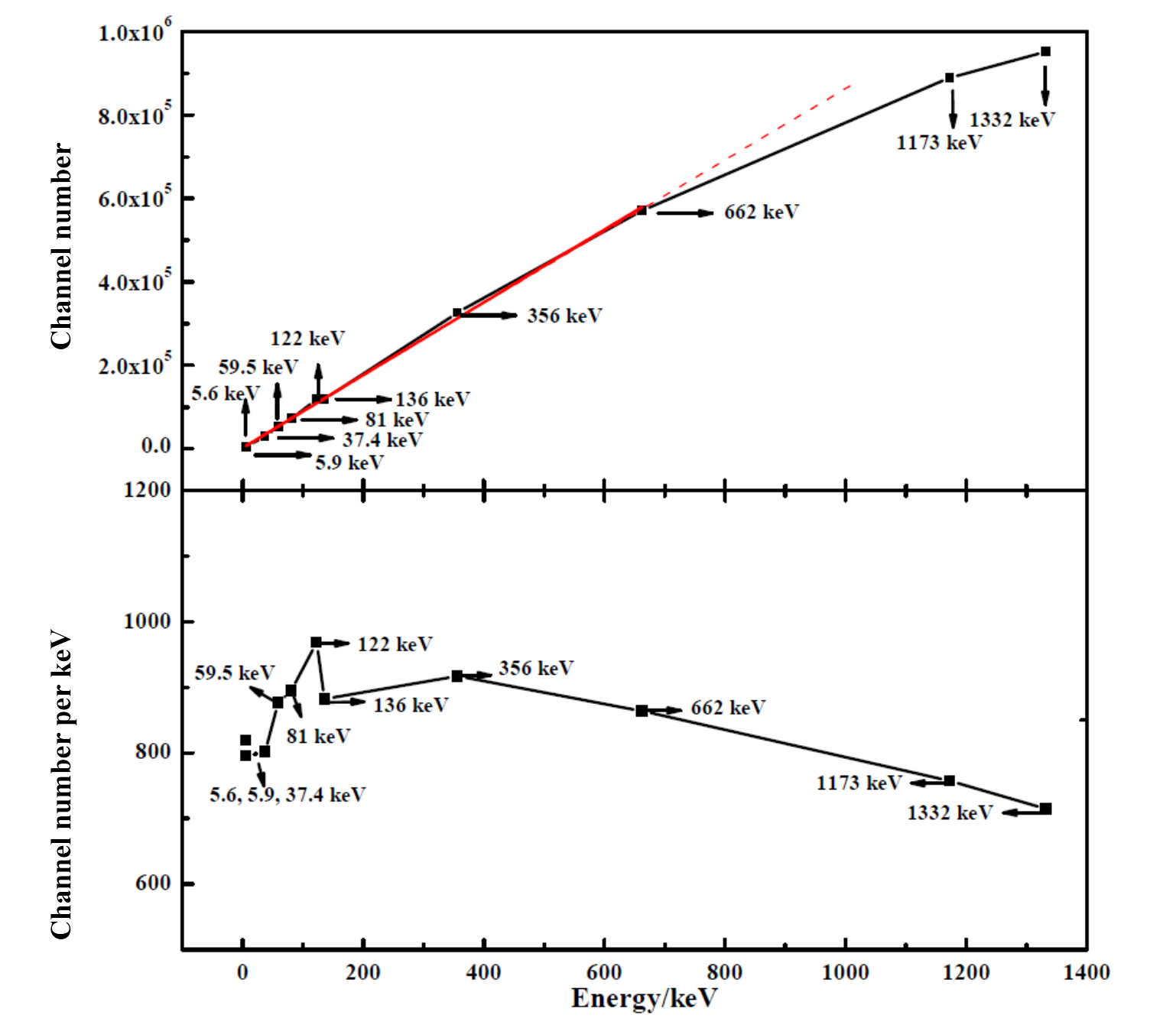}
	\caption{The plot of $E_{gamma}$ versus channel number and channel number per keV. Signal area (top panel) and signal area per unit of energy (bottom panel) versus gamma-ray energy. Saturation occurs above 700~keV.}
	\label{linearity}
\end{figure}

The linearity of $LaBr_3$:$Ce$ crystals is excellent compared to most other scintillators. There are two factors that may influence the linearity besides the crystal quality: the response of the SiPM and the electronics. The linearity of SiPMs have a good performance according to previous experiments~\cite{1748-0221-11-01-C01078, 1748-0221-10-08-P08013}. There are 620~microcells/$mm^2$ and more than $10^6$ pixels for each array, so it is not saturated within the GECAM detection range. We will clarify it carefully with simulation in the future. 
Fig.\ref{linearity} shows the signal area (in channels) and its reduced value (per keV) as a function of gamma-ray energy. The signal area was determined by waveform integration within 2.1~$\mu$s window. The linearity is preserved up to 700~keV, and then the saturation effect appears. The reason is that the dynamic range of the waveform digitizer DT5751 is only 1~V. For higher gamma-ray energies the signal amplitude exceeds the digitalization range and therefore its area is not
fully integrated. Herce, our next work is to design a special circuit, which is able to supply a high gain for low-energy rays and a low gain for high-energy rays to record all the intact waveforms.
\subsection{Uniformity}
\begin{figure}[H]
	\centering
	\includegraphics[width=8cm]{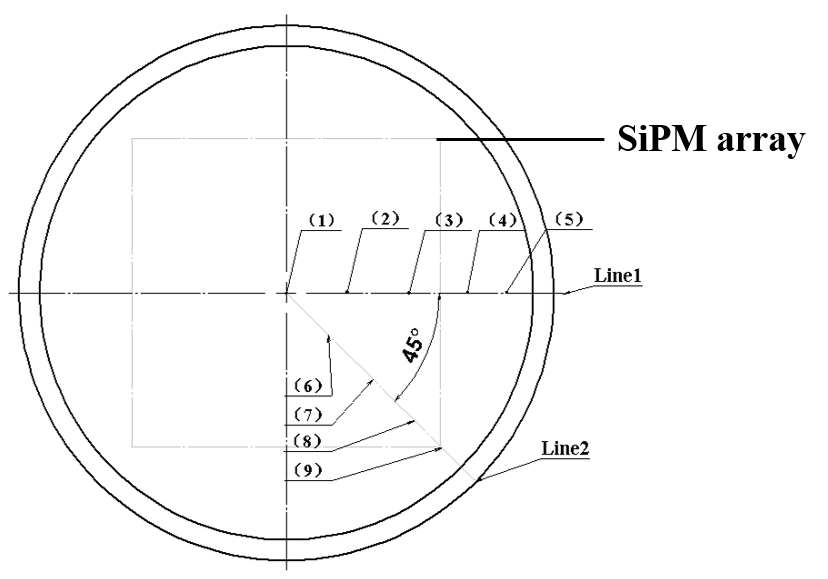}
	\caption{The placement of ${}^{241}Am$ source above the $LaBr_3$ crystal}
	\label{cricle}
\end{figure}

\begin{figure}[H]
	\centering
	\includegraphics[width=9cm]{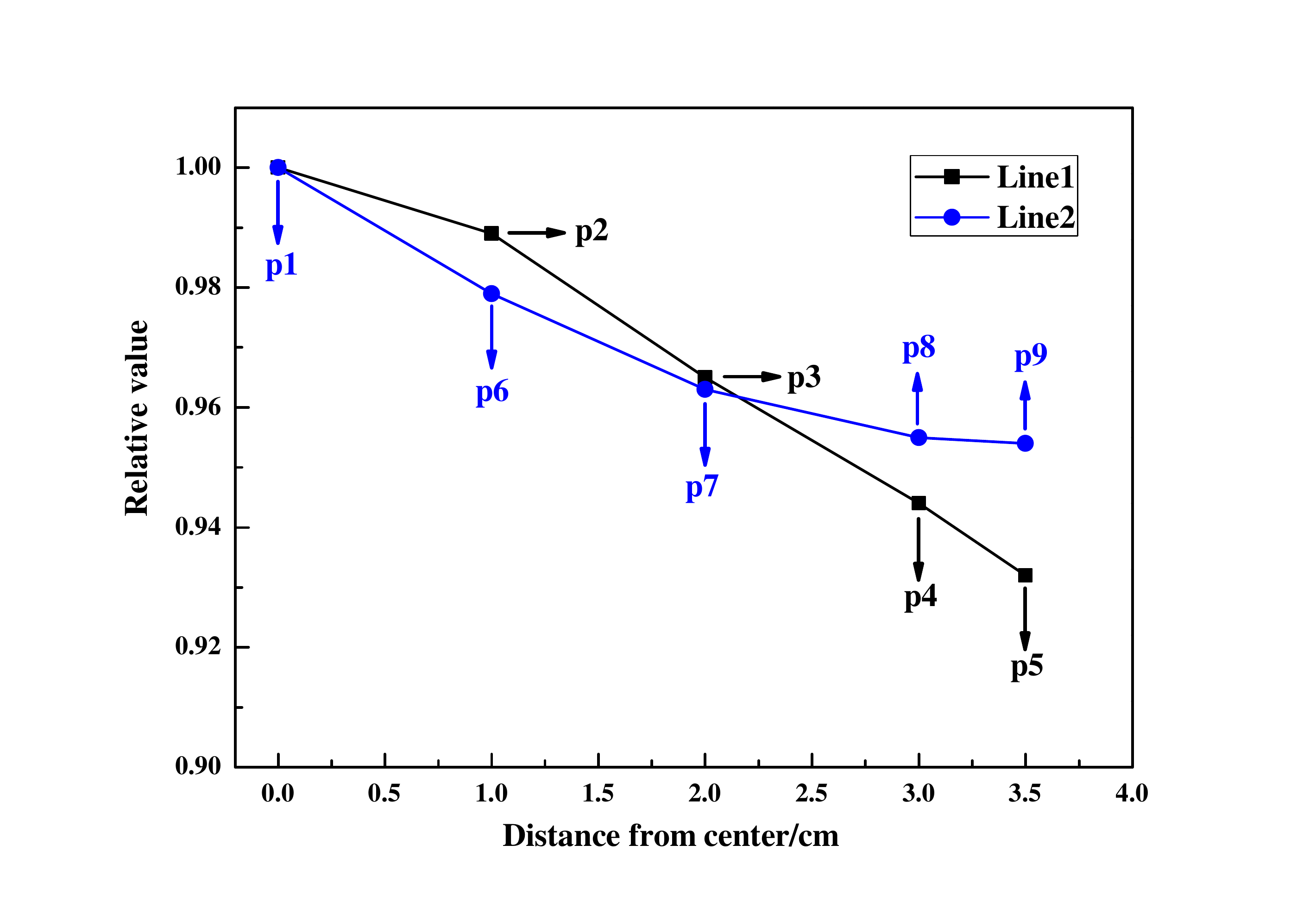}
	\caption{The Uniformity of the $LaBr_3$ crystal from center to edges.}
	\label{uniformity}
\end{figure}

The uniformity has been explored by using a collimated ${}^{241}Am$ 59.5~keV X-ray source. The source is positioned 25~mm away from the Beryllium window through a 3~mm diameter 20~mm thick lead collimation hole. A total of 9 points spaced by 1~mm were tested, as shown in Fig.\ref{cricle}, being the end points (5 and 9) 2~mm from the edge of the beryllium window. As shown in Fig.\ref{uniformity}, the vertical axis shows relative light collection efficiency assuming the efficiency at
the crystal center to be 1, and the efficiency declines $5\%$-$7\%$ as the location moves from the center to the edge.  The active area of the SiPM is $50.44\times50.44~mm^2$, and it cannot cover the crystal surface totally. Even though the Teflon film is used to increase reflectivity, the edge position lost more photons than the center. Our next work is to design a rounded SiPM array with the same size as the $LaBr_3$ crystal to completely cover the surface and obtain a better uniformity.
\section{Summary and Discussion}
GECAM is a dedicated space telescope that is designed for detecting high-energy electromagnetic counterpart. As its main detector, GRD is designed to be sensitive to X- and gamma-rays from 6~keV to 2~MeV. In the present paper, we reported on measurements of the GRD prototype performance. It is capable of detecting the X-rays as low as 5.6~keV, which satisfies the GECAM requirements. The energy resolutions are $6.5\%$ at 662~keV and $3.38\%$ at 1332~keV, and the linearity is preserved up to 700~keV due to the amplitude of high-energy ray exceeds the dynamic range of digitizer. The difference in uniformity between the center and the edges is about $7\%$, and the detection efficiency of the entire design is $70\%$ for 5.9~keV. Compare to the PMT-based similar detectors that are currently running in space like Fermi GBM~\cite{15} and CALET CGBM~\cite{16}, SiPM-based low-energy sensitive spatial gamma-ray detector is compact and innovative. 
\par\indent 
Our next work is to optimize the GRD design and update the circuit to enlarge the dynamic range, especially for the high-energy gamma-rays. A new rounded SiPM array is being designed to solve the problem of uniformity and improve the energy resolution. This new type of gamma-ray detector can be used in various field in the future.

\acknowledgments
The research is mainly supported by the Key Research Program of Frontier Sciences, Chinese Academy of Sciences (Grant NO. QYZDB-SSW-SLH012), and supported in part by the CAS Center for Excellence in Particle Physics (CCEPP), and supported in part by the National Natural Science Foundation of China( Grant No. 11775252),  and supported in part by the Young Scientists Fund of the National Natural Science Foundation of China (Grant No. 11305194 ).

\end{document}